# High-Throughput Information Storage in An Intelligent Response Phosphor

Dangli Gao,[a]* Zhigang Wang,[a] Xiangyu Zhang,[b] Qing Pang,[a] and Xiaojun Wang[c]*

[a] College of Science, Xi'an University of Architecture and Technology, Xi'an, Shaanxi 710055, China

[b] College of Science, Chang'an University, Xi'an, Shaanxi 710064, China

[c] Department of Physics, Georgia Southern University, Statesboro, GA 30460, USA

ABSTRACT

Persistent phosphor has emerged as a promising candidate for information storage due to the rapid accessibility and low-energy requirements. However, the low storage capacity has limited its practical application. Herein, we skillfully designed and developed $NaGdGeO_4:Pb^{2+},Tb^{3+}$ stimulated phosphor by trace doped $Sm^{3+}$. As expected, this phosphor demonstrates the larger carrier capacity than traditional commercial $SrAl_2O_4:Eu^{2+},Dy^{3+}$ phosphors and super-strong thermo-stimulated luminescence (TSL) that is three times greater than its photoluminescence (PL) intensity (PL efficiency: 17.3%). A mechanism of the enhanced and controllable TSL is proposed based on electron-hole defect pair structure. We further present a high-throughput optical data recording in five dimensions in a single fluorescent film recording layer. The findings described here provides not only a universal approach for construction TSL materials, but also a new paradigm for future generation optical storage technology.

## 1. INTRODUCTION

Information storage is essential in today's big-data era. In the light of the "Data Age 2025" report, by 2025, global data is expected to reach 163 ZB, equivalent to over 20000 GB per person.[1] Conventional hybrid hard disk storage technology is not up to the task due to its limitation in physical space and excessive energy consumption.[2] Optical data storage has risen as a standout contender for fourth-generation storage technologies in the context where the optical diffraction limit has been broken.[3-5] The new generation of optical storage technologies, such as optical quantum data storage, multidimensional storage technology, holographic technology, near-field/far-field optical technology, and integrated optical technology, are being intensively studied and have achieved many significant breakthroughs. However, suitable optical storage materials are very scarce.

Persistent storage phosphors hold a promising solution for their rapid accessibility and low-energy requirements. These materials can quickly record and read data by capturing charge carriers inside traps under UV light irradiation, and then release them into luminescence upon external stimulation, providing a convincing prospect for optical memory media.[6-9] Notably, BaFBr:$Eu^{2+}$, a representative photo-stimulated luminescence (PSL) phosphor, has acquired the greatest commercial success and widely used in imaging plates for computer radiography.[10] However, persistent materials as optical data storage media have been limited by their low capacity and short storage cycles. To increase information storage density, previous researchers have proposed some strategies including wavelength multiplexing, intensity multiplexing, and spatial dimension.[11-16] However, weak thermo/photo-stimulated luminescence (TSL/PSL) and afterglow interference often lead to signal crosstalk.

We know that the root cause of the uncontrollability of TSL/PSL signals lies in the uncontrollability of traps. A typical phosphor contains two types of active centers, named emission centers and trap centers.[17-19] The emission center determines the emission wavelength, while the trap center state (i.e. trap type, concentration, and depth) mainly affects intensity and duration of the persistent luminescence (PersL)/TSL/PSL signals.[20] The most widely accepted and efficient PersL/TSL/PSL processes are described based on valence band (VB) and conduction band (CB) models, including the hole transfer model,[21] and electron transfer model.[22] All PersL models agree that PersL generation is closely related to traps.

Herein, based on multi-cationic lattice characteristics of NaGdGeO$_4$ (NGGO), a novel NaGdGeO$_4$:Tb$^{3+}$ PersL storage material has been developed by the engineering trap structures via doping Pb$^{2+}$ and trace amounts of Sm$^{3+}$, demonstrates several excellent optical characteristics. Firstly, high density deep traps were obtained based on a construction of the more stable electron-hole defect pair structure, resulting in high-density charging capacity, strong and on-demand controlled TSL. Secondly, high TSL efficiency (>17.3%) due to the involvement of both CB and VB in remote transport of charge carriers. Finally, phosphor as an information storage medium holds the high storage density/capacity and long lifetime. As a result, the phosphors as optical data storage media demonstrate a multidimensional photo-data recording in a single layer. The types of traps in the PersL phosphors are determined using experimental and theoretical methods, including Hall-effect, electron paramagnetic resonance (EPR) spectra, X-ray photoelectron spectroscopy (XPS), partial density of states (PDOS), dynamic PL spectra and rate equations.

## 2. EXPERIMENTAL SECTION

**2.1. Chemicals.** High-purity chemicals were brought from Aladdin, including Na$_2$CO$_3$ (99%), PbO (99.9%), GeO$_2$ (99.999%), Re$_2$O$_3$ (99.99%) (Re=Sm, Eu, Gd, Tb, Dy, Ho, Er, Tm, Yb, Lu, Sc, Y, La, Ce, Pr and Nd), ZnO (99.99%), Ga$_2$O$_3$ (99.99%), Cr$_2$O$_3$ (99.99%), SrO and Al$_2$O$_3$ (AR).

**2.2. Sample preparation.** NaGdGeO$_4$:0.5 %Pb$^{2+}$,1.0 %Tb$^{3+}$,2.5*10$^{-6}$ % Re$^{3+}$ (NGGO:Pb,Tb,Re, Re=Sm, Eu, Dy, Ho, Er, Tm, Yb, Lu, Sc, Y, La, Ce, Pr and Nd) phosphor was prepared via a high-temperature solid state reaction approach using NaCO$_3$, GeO$_2$, Gd$_2$O$_3$, PbO, Tb$_4$O$_7$, Sm$_2$O$_3$, Eu$_2$O$_3$, Lu$_2$O$_3$, Dy$_2$O$_3$, Pr$_6$O$_{11}$, Er$_2$O$_3$, Nd$_2$O$_3$, Ho$_2$O$_3$, Tm$_2$O$_3$, Sc$_2$O$_3$, La$_2$O$_3$, Y$_2$O$_3$, Nd$_2$O$_3$, Yb$_2$O$_3$, and Tb$_4$O$_7$ as precursor materials. According to the stoichiometric ratio of NaGdGeO$_4$:x %Pb$^{2+}$,y %Tb$^{3+}$,2.5*10$^{-6}$ % Re$^{3+}$ (x = 0, 0.5 and y=0, 0.5, 1.0), the starting materials were prepared and mixed sufficiently in an agate mortar. Subsequently, the finely mixed powder was pre-calcined in a 950 °C chamber resistance furnace in the air for 6 h. Finally, the pre-annealed NGGO:Pb$^{2+}$,Tb$^{3+}$,Re$^{3+}$ powders were further sintered at 1100 °C for 6 h, and then naturally cooled to ambient temperature.

ZnGa$_2$GeO$_{12}$:Cr$^{3+}$ phosphors were prepared via high-temperature solid-state approach. During the synthesis process, first, weigh the raw material in the light of the stoichiometric ratio, and grind the weighed drug in an agate mortar for 30 min to obtain a uniform mixture of fine powder. Subsequently, the sample is placed in a high-temperature tubular sintering furnace for pre-calcination, raised to 1000 °C at a rate of 3.3 °C/min, and held for 5 h. After cooling, take out the sample and grind it for 30 min, raise it to 1200 °C at the same rate, and maintain it for 8 h. The phosphors were reground to obtain a final long afterglow sample.

SrAl$_2$O$_4$:0.5%Eu$^{2+}$,0.5%Dy$^{3+}$ phosphors were also synthesized using high temperature solid-state approach, and the initial materials were weighed and ground for 1 h, and then the mixture was calcined at 1400 °C for 5 h in Ar$_2$&H$_2$ atmosphere to obtain green commercial phosphors.

**2.3. Computational details.** The density functional theory calculation is conducted using projector augmented-wave approach, which is conducted on Vienna *ab initio* simulation package.[23,24] The Perdew-Burke-Ernzerhof functional with generalized gradient approximation (GGA) is employed for illustrating the exchange-correlation interaction.[25] A 1×2×3 NGGO supercell (containing 168 atoms) is adopted for simulating the doping process. The kinetic energy cutoff value for planar wave basis expansion is set to 400 eV, while the convergence standards of electron self-consistent energy and atomic force are set to $10^{-5}$ eV and 0.01 eV, respectively. The first Brillouin zone was sampled by a Γ-centered 2×2×2 k-mesh for structure relaxation, which then expanded to be 4×4×4 for accurate electronic properties calculations. To describe strongly correlated effect of localized 4f electrons, the GGA+$U$ scheme suggested by Dudarev *et al.*[26] is conducted where the tested Coulomb repulsion parameters $U$ for Gd and Tb are 7.5 eV and 3.0 eV respectively.

**2.4. Fabrication of flexible phosphor films.** The as-synthesized NGGO:Pb,Tb,Sm PersL phosphors were homogeneously mixed with a certain amount of silica gel and coagulant at a volume ratio of 1:1:1.[12] The mixture was carefully cast on a quartz chip mold to form a flat and smooth surface by throwing film method. The quartz chips are employed as mold to manage the size (20×22 $mm^2$) and thickness (~500 μm). Finally, after heating at 80 °C for 1 h under a slow heating rate in a blast drying oven, the flexible phosphor films were fabricated.

**2.5. Characterization.** The crystal structures of NGGO:$Pb^{2+}$,$Tb^{3+}$,$Re^{3+}$ PersL phosphors were characterized using an D/Max2400 X-ray diffractometer (XRD) under Cu Kα irradiation. A ZEISS Gemini 500 scanning electron microscopy (SEM) was employed for featuring the shape, size, and EDX spectra/mappings of NGGO:$Pb^{2+}$,$Tb^{3+}$ phosphors. An UV-Vis scanning spectrophotometer (Perkin-Elmer Lambda 35) with an integrating sphere in the range 200-900 nm was used for UV-Vis diffuse reflectance spectra of powder samples. The absorbance data can be acquired using converting reflectance data via Kubelka-Munk function. A Bruker A300 (70 K, 9.8536 GHz) was employed for collecting low-temperature EPR spectra. The X-ray photoelectron spectra (XPS) were recorded by a spectrometer (PHI 5600ci ESCA) equipped with monochromatized Al Ka radiation (1486.6 eV). Hall-effect measurements for the NGGO:Pb,Tb,Sm phosphor under UV excitation are carried out by using a LakeShore 8400 Hall instrument. For Hall measurement, phosphor samples were pressed and sintered in air into wafer (diameter: 8 mm, thickness: 0.98 mm), and two pairs of electrodes were fabricated on the surface using Au by vapor deposition method. The TL curves of samples were conducted using a home-made heating installation (the temperature range, 25–300 °C and the heating rate, 1 °C s$^{-1}$), locating in the sample chamber of the spectrometer (Horiba PTI). A Horiba spectrofluorometer (PTI QuantaMaster 8000) equipped with a xenon lamp (75 W) was used to investigate the spectral characteristic. In addition to a xenon lamp, two UV lamp (254 nm and 365 nm), and two NIR laser diodes (808 nm and 980 nm) were also used as light sources. The phosphor films were pre-irradiated for 3 min by the patterned photomask UV light (274 nm). The pictures of the optical information read-out were taken via a conventional Canon EOS 60D camera.

## 3. RESULTS AND DISCUSSION

NGGO host lattice that adopts an olivine-type arrangement, characterized by corner- or edge-shared polyhedra consisting of GdO$_6$, GeO$_4$, and NaO$_6$ units (Figure 1a), is developed. This structure provides three distinct cation sites: Gd$^{3+}$ (ionic radius r = 0.938 Å, coordination number (CN) = 6), Na$^+$ (r = 1.16 Å, CN = 6), and Ge$^{4+}$ (r = 0.53 Å, CN = 4). Given the comparable radius, valence, and coordination number, it is reasonable to expect that Tb$^{3+}$ (ionic radius r = 0.923 Å, CN = 6) as a luminescent center[27] or defect center[28] can effectively substitute for Gd$^{3+}$. While Pb$^{2+}$ is introduced as a maker of trap[29] due to its variable valence between divalent ions and tetravalent ions, accompanied with minute quantities of Re$^{3+}$ as a domino perturbation agent of crystal field environment.[30] Notably, the unit cell volume increases in doped NGGO phosphors relative to pure NGGO lattice due to substitution of $Pb^{2+}$ (r= 1.19 Å, CN= 6) for $Gd^{3+}$ (Table S1). XRD patterns of all NGGO phosphors with/without doping exhibit the same pure phase (JCPDS # 88-1776) (Figure 1b) in the orthorhombic space group Pnma. The refined pattern of NGGO:$Pb^{2+}$,$Tb^{3+}$,$Sm^{3+}$ shows excellent fitting results ($R_{wp}$ = 5.39% and $R_p$ = 4.48%) (Figure 1c), verifying the high phase purity of the samples. The spatial distribution of elements, as depicted in Figure 1d, along with corresponding energy-dispersive X-ray (EDX) analysis (Figure S1a), confirms the coexistence of Na, Gd, Ge, O, Pb, and Tb within NGGO:Pb,Tb,Sm crystals. Although $Sm^{3+}$ cannot be identified in the mapping due to trace amounts, the domino effect of $Sm^{3+}$ on the perturbation of crystal domain can be analyzed by spectroscopic methods in the subsequent discussion. Figure S1b demonstrates that NGGO:Pb,Tb,Sm phosphors is a broad bandgap (about 5.5 eV) materials.

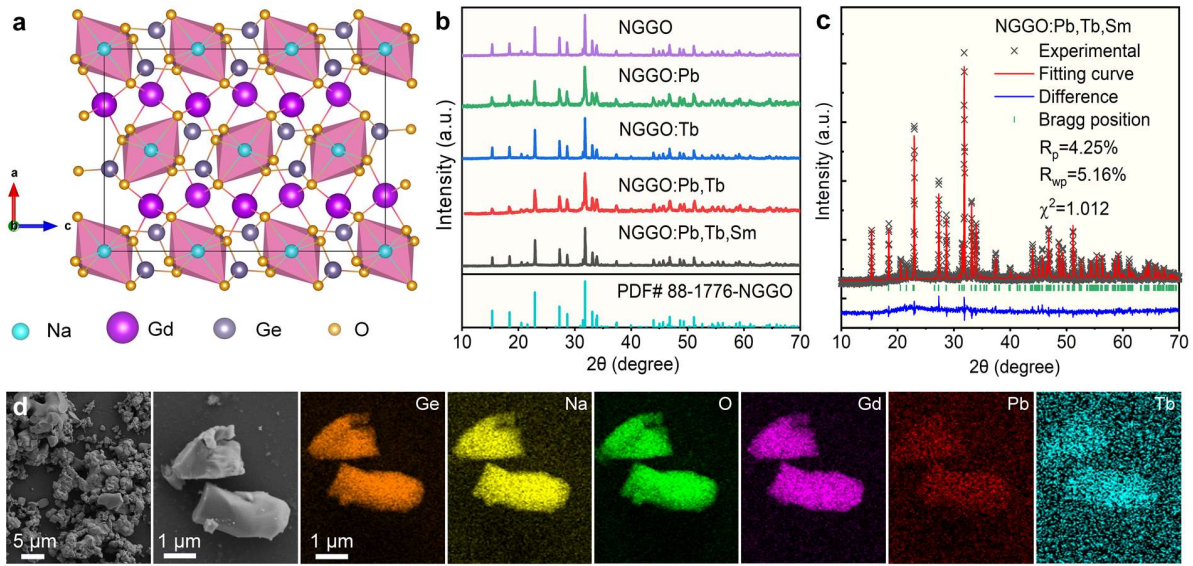

**Figure 1.** Crystal structure and EDX mapping of NGGO:Pb/Tb,Sm phosphor. (a) Orthorhombic NGGO crystal structure. (b) Typical XRD patterns of the NGGO:x%Pb,y%Tb,2.5*10$^{-6}$%Sm (x=0, 0.5 and y=0, 1.0) phosphors. The standard NGGO XRD data (JCPDS# 88-1776) is displayed as a reference at the bottom. (c) Rietveld refinement for the typical XRD pattern. (d) EDX mapping of NGGO:0.5%Pb,1.0%Tb, 2.5*10$^{-6}$ %Sm phosphor.

Figure 2 shows temperature dependent luminescence characteristics of Sm$^{3+}$ (2.5*10$^{-6}$ %Sm) trace doping NGGO:Pb,Tb phosphor. Upon excitation with 274 nm UV light, targeting the $^6P_J$ level of Gd$^{3+}$ in the phosphor, all PL emission spectra, originating from the Tb$^{3+}$: $^5D_{3,4} \rightarrow {}^7F_J$ (J = 6, 5, 4, 3) transitions (Figure 2a,b),[31] cover a wide spectral region from UV to visible light. No emissions from Gd$^{3+}$ are observed, indicating efficient energy migration/transfer from Gd$^{3+}$ to Tb$^{3+}$.[32] No Pb$^{2+}$ related transitions are observed even in NGGO:Pb singly doped phosphor. Monitoring host related transition at 400 nm in NGGO:Pb, the occurrence of Pb$^{2+}$ related transitions in the PL excited (PLE) spectrum of matrix related transitions indicate that it serves as a sensitizer (Figure S2a). After the Sm$^{3+}$ enter into NGGO:Pb, the related transitions of Pb$^{2+}$ almost disappeared, while the ground state to $^6P_{7/2}$ transition of Gd$^{3+}$ increased. This may be due to the doping of Sm$^{3+}$ causing Pb$^{2+}$ to become a new defect rather than a sensitizer. PLE spectra ($\lambda_{mon}$= 553 nm) originating from Tb$^{3+}$ ions reveal distinct sharp line peaks superimposed upon a broad band spanning from 200 to 500 nm (Figure 2a, Figure S2b).

Besides the PLE peaks of Tb$^{3+}$ (including 4f→5d transition peaking at 261 nm, $^7F_6 \rightarrow {}^5G_{2,3,4}$ at 337-400 nm and $^7F_6 \rightarrow {}^5D_4$ peaking at 481 nm f-f transitions), the sharp peaks from Gd$^{3+}$: $^8S_{7/2} \rightarrow {}^6P_J$ at 274 and 312 nm on PLE spectra can be clearly identified (Figure 2a and Figure S2b). Compared to PLE spectra of NGGO:Tb and NGGO:Pb,Tb, PL intensity assigned to 4f→5d transition of Tb$^{3+}$ at 261 nm in NGGO:Pb,Tb,Sm phosphor decrease and shift to right (Figure S2b), which may be due to downward shift of the bottom of conduction band (CB). Figure 2c shows temperature dependent PersL/TSL characteristic of NGGO:Pb,Tb, Sm$^{3+}$ charging at room temperature. A set of TSL emission bands (Figure 2c) is similar to PL peaks, assigned to the same luminescent center. As the temperature increases, the TSL signals increase and reach a maximum value at 140 °C

(Figure 2c).

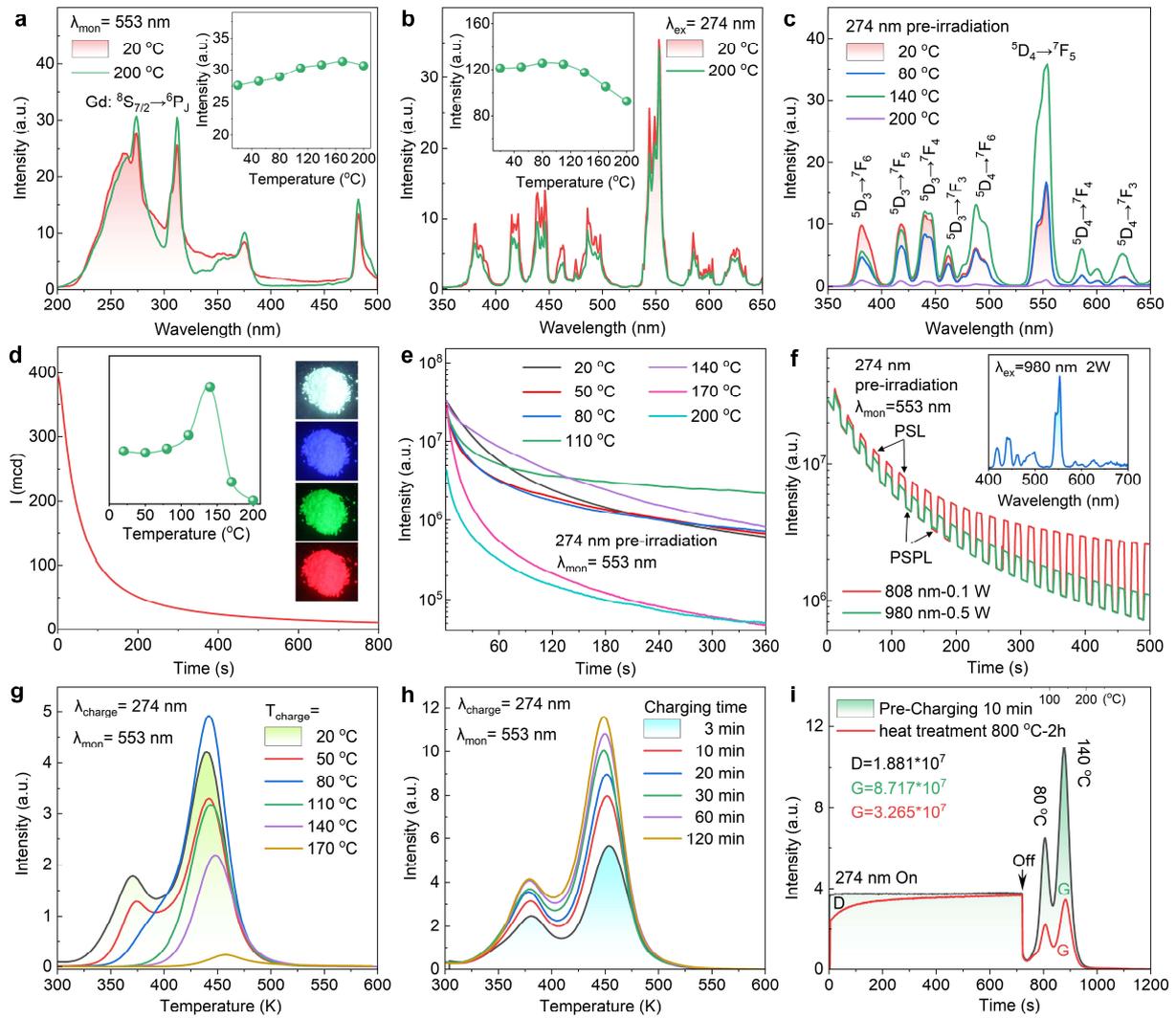

**Figure 2.** Luminescence properties of NGGO:Pb,Tb,Sm phosphor. (a) Temperature dependent PLE spectra ($\lambda_{mon}$= 553 nm) with the PL peak intensity (at 274 nm) shown in the inset as a function of temperature. (b) Temperature dependent PL spectra ($\lambda_{ex}$=274 nm) alongside the PL peak intensity (at 553 nm) as a function of temperature in the inset. (c) Temperature dependent PersL spectra charging at room temperature. (d) Absolute PersL intensity, with the PersL/TSL intensity plotted against temperature, and corresponding white-color PersL and tricolor PersL photos (under suitable filter) in the insets. (e) Temperature dependent PersL decay curves ($\lambda_{mon}$= 553 nm). (f)

Pulsed PSL and PSPL acquired from a decay curve under 980 nm (0.5 W) or 808 nm (0.1 W) laser stimulation (10/10 s). The inset is a PSL emission spectra. (g) TL curves charging at varying temperatures. (h) Charging time dependent TL curves. (i) Initial trap filling degree dependent PL (RT)-TSL (RT-250 °C) dynamic processes ($\lambda_{mon}$=553 nm) (Note that one phosphor had their traps cleared at 800 °C for 2 h for acquiring red measurement line and another phosphor had their traps filled by UV pre-irradiation for 10 min for the black line). Prior to the measurement of PersL, decay curves, the phosphors were pre-irradiated using a 274 nm UV lamp for 3 min.

The PersL intensity (initial intensity: 400 mcd, PL quantum efficiency: 17.3%) diminishes rapidly with time delay (Figure 2d). Notably, the intensity and duration of PersL/TSL can be finely modulated by temperature management in the inset in Figure 2d and 2e, with the TSL peak achieved at 140 °C, indicating an anomalous thermal quenching behavior. While PL intensity remained almost unchanged within the temperature range studied (insets in Figure 2a,b). Beyond PersL, slightly enhanced PSL and PSPL signals originating from $Tb^{3+}$ are discernible and can be successively illuminated under and after NIR laser irradiation (Figure 2f).

Figure 2g shows the TL curves of different charging temperatures. It is noted that the peak value of TL curve charging at room temperature is about 140 °C, which explains why the optimal TL emission spectrum at 140 °C in Figure 2c. The trap filling capacity (i.e., the TL spectrum area in Figure 2g) first increases and then decreases with increasing charging temperature, reaching a maximum at 80 °C, indicating a PersL anomalous quenching phenomenon. Figure S3 shows a comparison of TL creation spectra charging at the two different temperatures (i.e., RT and 80 °C). All TL profiles exhibit identical glow peak I and II after RT charging. The average depth (E) of Trap is estimated using trap depth formula $E = (0.94\text{Ln}\beta+30.09) \times kT_m$,[33] where $\beta$, $k$ and $T_m$ are the heating rate, the Boltzmann constant and temperature of the TL peaks, yielding values of 0.83 eV and 1.07 eV for Trap I and Trap II, respectively. When charging temperature is elevated to 80 °C from RT, Trap I almost cleared by heating, while the filling capacity of Trap II increases, reaching twice RT capacity (Figure 2g and Figure S3c). Note that the profiles of the two TL creation spectra are similar to the PLE spectra, but the TL creation spectra charged at 80 °C yield higher TSL intensity than those charged at RT (Figure S3).

Considering optical information storage, carrier storage capacity (i.e. trap filling capacity) is another important parameter. Trap filling capacity reaches saturation (i.e., the maximum value of the TL integral intensity) after UV charging for 2 h, demonstrating an exceptional storage capacity akin to a small battery (Figure 2h and S4a). This super high charge carrier storage capacity (Figure 2h) allows TSL intensity to be effectively regulated by charging wavelength (Figure S3), charging temperature (Figure 2g) and charging time (Figure 2h and Figure S5). In addition, the phosphor can also be charged using X-ray and common polychromatic light sources including sunlight and composite light containing extremely weak UV light alongside robust NIR light from a filtered xenon lamp emission (Figure S4b-d).

We next examine the transient PL-TSL processes of NGGO:Pb,Tb,Sm with and without pre-charging using UV light (Figure 2i). Comparison of PL-TSL curve characteristics of the two samples (including a sample pre-charged for 10 min and the other is completely cleared traps) shows that the PL process of pre-charged sample rapidly reaches a dynamic equilibrium, whereas in the cleared phosphor, PL increases gradually and then reaches the similar equilibrium under UV light irradiation. The disparity, denoted as 'D' (i.e., charging area),[34,35] between the two PL curves (shown as a shaded region) signifies the energy stored in the traps during the PL process. Turning off the UV light and heating simultaneously from RT to 250 °C, the pre-charged sample exhibits more than double trap filling capacity (*i.e.*, integral intensity in the TL curve area 'G') of the cleared sample. Additionally, the maximum TSL intensity acquired at 140 °C (Figure 2i) is four times higher than its balance intensity of PL (PL efficiency: 17.3% and initial PersL intensity of 400 mcd in Figure 2d). These results indicate that even after reaching a dynamic equilibrium in PL, the traps continue to capture charge carriers from the VB/CB (Figure 2i), leading to super high charge storage capacity and TSL intensity (Figure 2h,i). A complete charge-discharge and TSL process is shown in Video 1.

Figure 3a,b demonstrate the relationship between PL and TSL by the transient PL-TSL process via regulating the charging wavelength and temperature. When exciting $^6P_J$ level of $Gd^{3+}$ using 274/312 nm UV light, higher PL intensity is acquired relative to direct excitation of $Tb^{3+}$ at 261 nm, demonstrating effective energy transfer from Gd to Tb. Conversely, direct excitation of $Tb^{3+}$ leads to a larger charging area 'D' in the PL-charging process, indicating that $Tb^{3+}$ may act as a trap itself.

In order to further confirm the strong charging capacity and TSL ability of the NGGO:Pb,Tb,Sm phosphor, Figure 3c,d shows the PL-TSL process of commercial red $ZnGa_2GeO_{12}:Cr^{3+}$ and green $SrAl_2O_4:Eu^{2+},Dy^{3+}$ phosphors, where PersL/TSL intensity is much lower than PL intensity due to small carrier storage capacity. Compared with PL-TL of commercial $SrAl_2O_4:Eu^{2+},Dy^{3+}$ and red $ZnGa_2GeO_{12}:Cr$ phosphors in Figure 3c,d, NGGO:Pb,Tb,Sm phosphors in Figure 2i and Figure 3e exhibit much higher carrier capacity and stronger TSL. The trap filling capacity of NGGO:Pb,Tb,Sm phosphor is eight time bigger than that of $SrAl_2O_4:Eu^{2+},Dy^{3+}$. In addition to its ultra-high carrier capacity and strong TSL, NGGO:Pb,Tb,Sm phosphor also demonstrates the non-volatile information storage capability. Time delayed TL curves are a key technical means to reveal whether afterglow materials are suitable for non-volatile information storage. The time delayed TL curves (Figure 3e) indicate that at 1 h after turning off the light source, Trap II has an energy conservation rate of almost 100%. For 10-20 days after stopping excitation, the energy storage rate of Trap II decreases to about 50%, and thereafter the energy is almost frozen in the deep trap and cannot escape at RT. Figure S4b exhibits the strong TSL emission spectra acquired on 3 min and the fifth day after cessation of excitation.

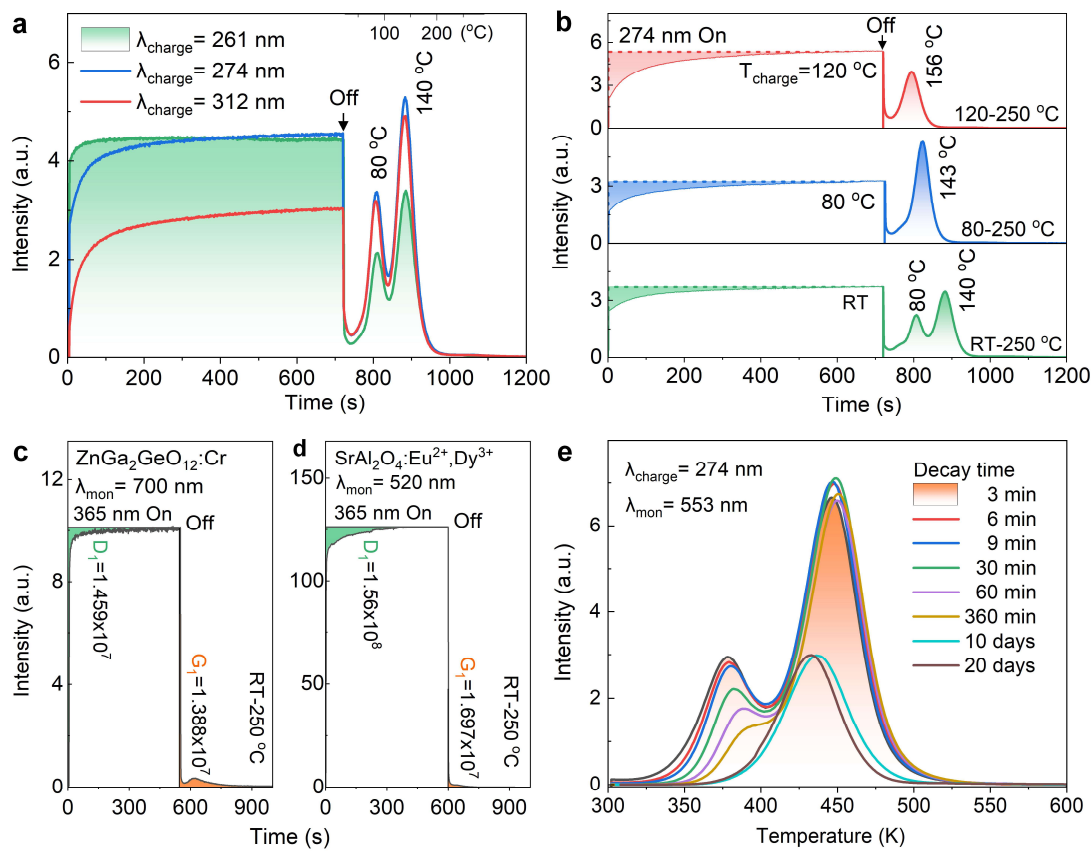

**Figure 3.** Transient PL-TSL processes of NGGO:Pb,Tb,Sm, ZnGa$_2$GeO$_{12}$:Cr and SrAl$_2$O$_4$:Eu,Dy phosphors, and delay time dependent TL spectra of NGGO:Pb,Tb,Sm. (a) Charging wavelength dependent PL (RT)-TSL(RT-250 °C) of NGGO:Pb,Tb,Sm. (b) Charging temperature dependent PL (charging temperature)-TSL (charging temperature-250 °C) of NGGO:Pb,Tb,Sm. (c,d) PL (RT)-TSL (RT-250 °C) dynamic process of ZnGa$_2$GeO$_{12}$:Cr[17] and SrAl$_2$O$_4$:Eu,Dy phosphors.[21] (e) Delay time dependent TL spectra of phosphor pre-charging with 274 nm at RT for 3 min. Note that before measuring, the sample was treated at 800 °C for 2 h to completely clear the trap.

To shed more light on the mechanism of the super-strong TSL, we firstly validate the role of trace $Re^{3+}$ doping in TSL process. A series of NGGO:Pb,Tb samples doped with trace amounts of $Re^{3+}$ was studied. All the samples exhibit similar TL profiles (Figure S6c, Figure S7 and S8) and enhanced TL integral intensity (Figure 4a). Among them, NGGO:0.5%Pb,1.0%Tb,Sm phosphor stands out with a significantly enhanced TSL intensity compared to other samples (Figure 4a and Figure S6-S8) and a six-fold increase in TL integral intensity relative to NGGO:0.5%Pb,1.0%Tb. However, all the samples doped with trace $Re^{3+}$ show a slight decrease in PL intensity, as shown in the bottom panel of Figure 4a, indirectly supporting the increase in trap density. Effective PersL energy transfer from $Gd^{3+}$ to $Tb^{3+}$ was confirmed by examining the significant reduction in TSL intensity of $Tb^{3+}$ with the decrease in concentration of $Gd^{3+}$ (Figure S8d),[27,31] where partial $Gd^{3+}$ are substituted with the optically inert $Lu^{3+}$.

In an effort to further confirm the type of trap in NGGO:Pb,Tb,Sm phosphor, the PL-TSL dynamic processes, TL spectra, electron paramagnetic resonance (EPR), X-ray photoelectron spectroscopy (XPS), Hall effect and partial density of states (PDOS) analyses are conducted. Figure 4b shows a comparison of the PL-TSL dynamic processes for pre-charged and completely emptied NGGO:Tb, NGGO:Pb,Tb, and NGGO:Pb,Tb,Sm under UV irradiation at RT. The PL intensity shows a trend of NGGO:Tb > NGGO:Pb,Tb = NGGO:Pb,Tb,Sm, while the TL curve integrated intensity of TL curve has an order of NGGO:Tb < NGGO:Pb,Tb < NGGO:Pb,Tb,Sm (Figure 4b). The optimal TSL intensity is about three times greater than its PL intensity at 140 °C, along with a larger integrated area of TL curve in NGGO:Pb,Tb,Sm than in NGGO:Tb phosphor (the bottom panel in Figure 4b), indicating that doping of $Pb^{2+}$ and $Sm^{3+}$ greatly increases the density of trap II. In Figure 4c,e and S8, all $Tb^{3+}$ doped samples, including NGGO:Tb, NGGO:Pb,Tb and NGGO:Pb,Tb,Re, exhibit similar glow peak I and II, a departure from TL peaks observed in NGGO and NGGO:Pb (Figure 4c) as well as all samples prepared under reducing atmospheres (Figure 4d). Notably, integral TL glow peak II area (i.e., trap density or filling capacity) in doping trace $Re^{3+}$ samples is much larger than their counterparts (Figure 4b and Figure S9). Figure 4e demonstrates that $Sm^{3+}$ doping has a greater impact on NGGO:Pb relative to NGGO:Tb. NGGO:Pb,Tb, NGGO:Tb and NGGO samples obtained under reducing atmospheres display a sharp decline on TSL intensity (Figure 4d) relative to in air (Figure 4c), and trap II on TL spectra in NGGO:Tb and NGGO:Pb,Tb phosphors nearly vanishes in an anaerobic environment indicating that trap II is not oxygen vacancies. The EPR signals have not changed in NGGO:Tb, NGGO:Pb,Tb and NGGO:Pb,Tb,Sm samples before and after UV *in situ* irradiation (Figure S10), suggesting that trap I and II are not oxygen vacancies.

Figure 4f illustrates the XPS spectra of the O 1s state for the NGGO:Pb,Tb phosphor before and after trace doping $Sm^{3+}$. Based on the fitting results, there are three different peaks located at 529.60 eV-532.15 eV, which are associated with the lattice oxygen $O_I$ (529.60), $O_{II}$ (i.e., one $O_I$ is connected to two Na, a Ge and a Gd, and another $O_{II}$ is connected to two Gd, a Na and a Ge) and absorbed oxygen (532.15 eV), respectively.[35,36] High-resolution XPS analysis of the $O_{II}$ 1s peak (Figure 4f) reveals a discernible shift of 0.31 eV towards lower binding energy after trace $Sm^{3+}$ doping, indicating the existence of $O_{II}$ with lower valence and Pb/Tb with higher valence. All high-resolution XPS spectra of Pb 4f in Figure 4g,h exhibit double-peak characteristics, indicating the coexistence of the $Pb^{2+}$ and $Pb^{4+}$. While after doping $Sm^{3+}$ in NGGO:Pb,Tb phosphor (Figure 4g), the proportion of $Pb^{4+}$ defect state intensity (peaking at 144 eV) increased (red line) relative to black line. Before and after charging, there is no obvious change on the ratio of $Pb^{2+}$ to $Pb^{4+}$ in NGGO:Pb,Tb,Sm (Figure 4h).

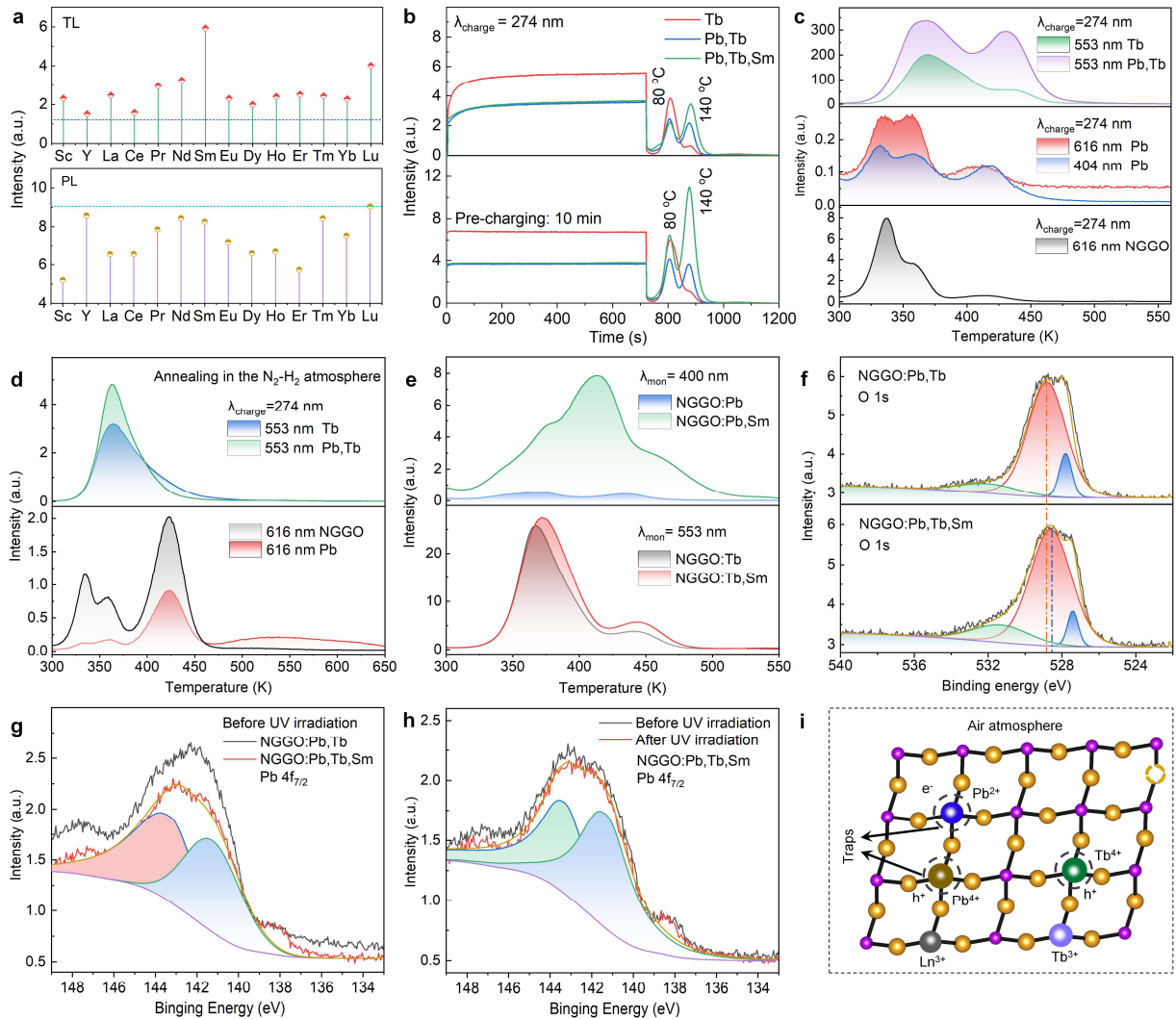

**Figure 4.** Effect of doping on TL spectra and defect structures of NGGO Phosphors. (a) Influence of trace $Re^{3+}$ doping on integral intensity of TL spectra (data derived from Figure S8) and PL intensity of NGGO:Pb,Tb phosphors (data derived from Figure S9), with a TL integral intensity/PL intensity (blue/green horizontal lines) of NGGO:Pb,Tb as a reference. (b) Comparison of PL-TSL dynamic processes for completely emptied and pre-charged NGGO:Tb, NGGO:Pb,Tb and NGGO:Pb,Tb,Sm samples monitored at 553 nm. (c,d) Influence of calcination atmosphere (air for samples in c and $N_2$:$H_2$=3:1 reduction atmosphere for samples in d on TL curves. (e) Influence of trace $Sm^{3+}$ doping on TL curves of NGGO:Pb and NGGO:Tb. (f,g) Influence of trace $Sm^{3+}$ doping

on O 1s and Pb 4f high resolution XPS spectra of NGGO:Pb,Tb. (h) Pb 4f high resolution XPS analysis of NGGO:Pb,Tb,Sm phosphor before and after charging, and (i) Possible electron-hole pair defect-structure diagram in air. Preceding TL curve measurements, the samples were pre-irradiated under UV light for 3 min.

The PDOS analysis underscores the predominant presence of O 2p and Gd:5d,4f levels in both VB and CB across all models, suggesting a robust hybridization effect between Gd 5d,4f levels and O 2p levels (Figure S11). Upon introduction of $Pb^{2+}$ and $Tb^{3+}$ into the host lattice, two defect bands emerge within the band gap. One of these bands is a hybrid combination of Pb 6s and O 2p, while the other stems from the Tb 4f states. Hall effect measurements at RT confirm that carrier type is hole and VB is involved in the transport of charge carriers (Table S2). Based on time-dependent TL measurements, released charge carriers of the RT phosphor are mainly shallow trap I, since charge carrier in deep traps are generally not released at RT (Figure 4a). Therefore, we can deduce that shallow trap I is $Tb^{3+}$ with a hole. Considering the measurement results above together, upon the introduction of $Pb^{2+}$ into the lattice, defect reaction $2Pb^{2+} \xrightarrow{2Gd^{3+}} Pb^{2'+}_{Gd^{3+}} + Pb^{4\bullet+}_{Gd^{3+}}$ may occur (Figure 4i) to maintain electrical neutrality.[37-40] Trace $Re^{3+}$ doping gives rise to the formation of $Pb^{4+}$ (Figure 4e), and then a formation of $Pb^{4+}$ (as an electron trap)-$Pb^{2+}/Tb^{3+}$ (as a hole trap) electron-hole pairs. This also accounts for the disappearance of trap II in a reducing atmosphere (Figure 4d), where $Pb^{4+}$ defects are absent (Figure 4h). Taken together, these findings suggest that trap II is likely closely associated with $Pb^{4+}$-$Pb^{2+}$ ($Tb^{3+}$) electron-hole pair defect cluster (Figure 4i).[38,41] $Pb^{4+}$ defect density can be improved by doping trace $Re^{3+}$ (Figure 4a) for lattice distortion[42,43] or elevating annealing temperature (Figure S12). In fact, the self-redox reactions of $Mn^{4+}$ and $Eu^{3+}$, or both holes and electrons of $Bi^{2+}$ have been reported.[44-46]

A TSL mechanism is proposed based on hole-electron defect pairs, as shown in Figure 5. $Gd^{3+}/Tb^{3+}$ ground state electrons are initially elevated from to the excited states upon UV lamp irradiation (process 1) and then electrons ionize to the CB or hole move to VB (process 2). Then, electrons/holes migrate within CB/VB and are captured by traps (process 3), or recombine with Tb/Gd or defect centers. Finally, the electrons/holes migrate in the CB/VB until they are captured (process 3) by Tb/Gd (leading to PL) or defect centers (traps harvest energy). Under photo/thermo stimulation, the electrons trapped by traps are discharged and then recombined with excited states of Gd/Tb (process 4,5), resulting in TSL/PSL of $Tb^{3+}$ (process 6). Higher temperatures help promote thermoelectric ionization (process 2) and energy transfer/transfer efficiency (process 5) in Figure 5, which may cause the anomalous thermal TSL effect. The NGGO host contains essential $Gd^{3+}$ ions that facilitate energy transfer from traps to $Tb^{3+}$ while minimizing energy loss through direct energy transfer from Tb to defects. The electron-hole pair mechanism of the variable valence elements suggested here is also suitable for explaining the afterglow process of previous reported $SrAl_2O_4:Eu^{2+}$, $SrAl_2O_4:Eu^{2+},Dy^{3+}$ and $Sr_2MgSi_2O_7:Eu^{2+},Dy^{3+}$ phosphors.[30-33]

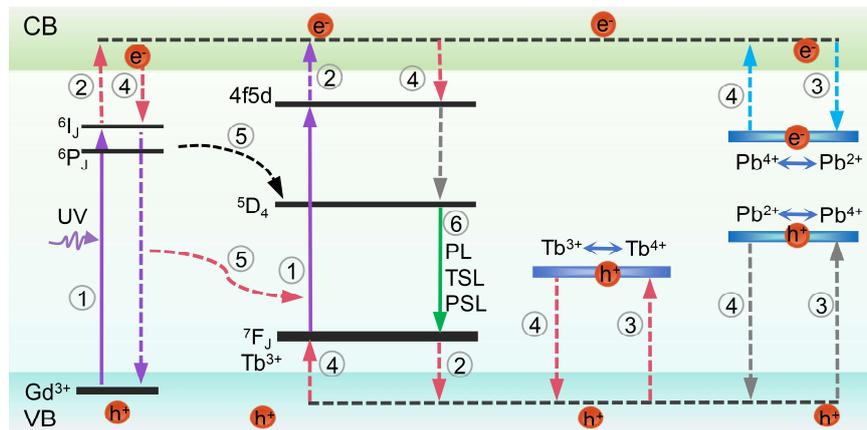

**Figure 5.** Energy storage and TSL mechanism of NGGO:Pb,Tb,Sm phosphor.

The successful control of trapping and de-trapping carriers based on traps' intelligent response to temperature in single phosphor prompted us to design and fabricate luminescent thin film as a storage medium (Figure S13), enabling the storage of multiple individually addressable patterns within a single recording layer (Figure 6).[2,3,47,48] Figure 6a illustrates the encoding and retrieval of information in specific depths of traps by controlling temperature and time. Using mask-patterned 274 nm UV light irradiation, distinct patterns such as the three cloud ID logo, including "Sending" (blue), "Transportation" (red) and "Receiving" (green), are written into varying depths of traps at different temperatures (Figure 6a). The filling capacity of traps is accurately controlled by managing charging time—1 min, 2 min, and 7 min. As depicted in Figure 6b, "Receiving" is initially written into the deepest trap at 110 °C, followed by "Transportation" in the second trap at 80 °C, and finally "Sending" in the shallowest trap. These stored patterns can only be retrieved when the phosphor film receives adequate activation energy.[15] Based on the intelligent response of trap to temperature, the sequential, individual, and distinct read-out of the pre-stored "Sending", "Transportation", and "Receiving" patterns are demonstrated in Figure 6b under thermo-stimulation at designated temperatures like 60, 100 and 170 °C. Subsequently, these glowed patterns are captured using a cameras or mobile phones. Notably, the overlapping pattern of the third layer signal, with distinct grayscale values, functions as an information encryption layer (i.e., the behavior of traditional afterglow phosphors). When heated to 170 °C, this layer is instantaneously recorded, followed by the actual image of the "Receiving". Consequently, diverse information can be stored and retrieved from the same location, achieving multidimensional optical data storage.

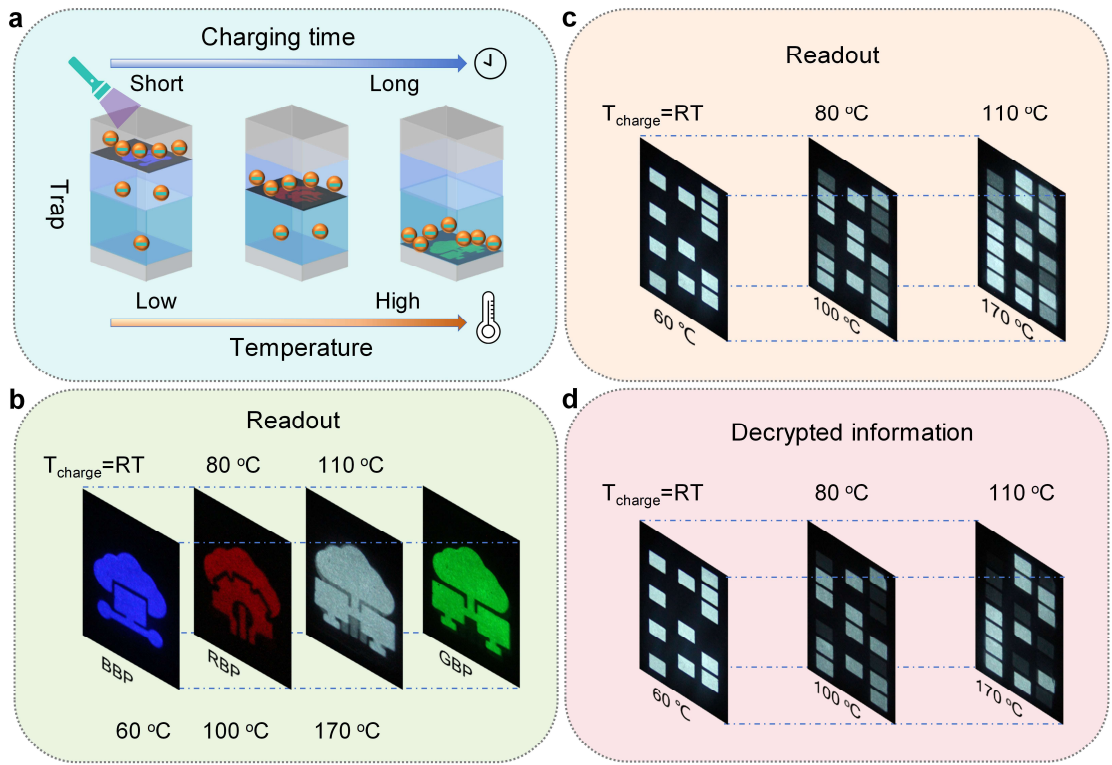

**Figure 6.** Multidimensional optical data storage read-out and encryption using NGGO:Pb,Tb,Sm flexible phosphor film. (a) Schematic representation of multilevel optical data storage write-in through selective filling of traps, controlled by varying charging temperature and time. Information is written using 274 nm UV light irradiation for 1 min, 2 min, and 7 min to fill shallow, medium, and deep traps. (b) Read-out of ID logo including "Sending" (blue), "Transportation" (red) and "Receiving" (green) stored in three-layer traps with a suitable filter. The first, second and fourth images depict the read-out of ID logo, while the third image represents the set key. (c) Read-out of binary codes. The second image shows the combination of the first and second images, while the third image displays the combination of the first, second and third images. (d) Independent and orthogonal read-out of information written in each layer of traps for binary codes. All images are 20 × 22 mm$^2$ in size. Imaging parameters: Manual mode, ISO 3200, 4-second exposure. BBP, RBP and GBP are blue, red, and green bandpass filters in b.

Similarly, three distinct binary codes are encoded into the traps of a layer using mask-patterned UV light irradiation, and then the luminescent patterns of these binary codes are retrieved using a camera with thermal assistance. Figure 6c,d shows the process where the information of the first card is accessed upon heating to 60 °C. Elevating the temperature to 100 °C allows for the selection of both the combined signal of the first and second cards with different grayscale values (Figure 6c) and the individual signal of the second card (Figure 6d) to be recorded during heating at 110 °C. Similarly, the combined signal of the first, second, and third cards (Figure 6c) and the individual signal of the third card (Figure 6d) can also be selectively retrieved. Consequently, combined both the TSL intensity grayscale values and trap domains, the three-trap-state multiplexing effectively offers a total of five multiplexed states within one recording layer. While the existing multiplexing techniques are orthogonal in the three-trap state, the two additional layers with overlapping segments are independent and often employ varying read-out temperatures to selectively erase undesired information from other layers. It is essential to highlight that simultaneous information read-out and intricate grayscale or color encryption can be achieved after UV or sunlight charging (Figure 6d). For instance, adaptable charging temperatures, reading temperatures, and charging durations enable the creation of multiple encrypted patterns with distinct grayscale values for comparison (Figure 6c). Additionally, multilevel encryption can also be achieved through simple photo/thermo stimulation after UV/solar irradiation (Figure S14).[49]

## 4. CONCLUSIONS

In summary, the NGGO:Pb,Tb,Sm PersL phosphors have been developed and they demonstrate super high capacity of loading energy carriers, which allows them to store five-dimensional optical recording within a physical single-layer storage surface by controlling trap loading and releasing on demand, boasting high-throughput optical data storage. TSL mechanism is proposed based on electron-hole defect pairs, which improves the PersL mechanism and expands the range of potential applications of PersL materials. Considering the advantage and versatility offered by these materials, the findings have important implications in advanced information storage, encryption and visual temperature sensing. Given the substantial flexibility of phosphors, our discovery may also stimulate other new applications and design the concepts on PersL materials.

## ASSOCIATED CONTENT

**Supporting Information**

Refined structural parameters, EDX and UV-Vis absorption spectra, PLE, PL, PersL spectra, PersL decay curves, TL curves, PL-TSL dynamic processes, effects of $Re^{3+}$ micro-doping and $Gd^{3+}$ concentration on TL, EPR spectroscopy, density functional theory calculations, Hall effect measurement, multilevel read-out and encryption from a flexible phosphor film (PDF).

## AUTHOR INFORMATION

**Corresponding Authors**

**Dangli Gao** − *College of Science, Xi'an University of Architecture and Technology, Xi'an, Shaanxi 710055, China*; orcid.org/0000-0001-5907-6817; Emails: gaodangli@163.com, gaodangli@xauat.edu.cn.

**Xiaojun Wang** − *Department of Physics, Georgia Southern University, Statesboro, GA 30460, USA*; orcid.org/0000-0003-1506-0762; Email: xwang@georgiasouthern.edu.

**Notes**

The authors declare no competing financial interest.

## ACKNOWLEDGMENT

This work was supported by the National Natural Science Foundation of China (No. 11604253), Shaanxi Fundamental Science Research Project for Mathematics and Physics (No. 23JSY003) and Shaanxi Key Science and Technology Innovation Team Project (No. 2022TD-34).

**REFERENCES**

(1) Hilbert, M.; López, P. The world's technological capacity to store, communicate, and compute information, *Science* **2011**, *332*, 60.

(2) Reinsel, D.; Gantz, J. Data Age 2025: The evolution of data to life-critical don't focus on big data, *Framingham: IDC Analyze the Future* **2017**.

(3) Zhao, M.; Wen, J.; Hu, Q.; Wei, X.; Zhong, Y.; Ruan, H.; Gu, M. A 3D nanoscale optical disk memory with petabit capacity, *Nature* **2024**, *626*, 772.

(4) Xiong, B.; Liu, Y.; Xu, Y.; Deng, L.; Chen, C. W.; Wang, J. N.; Peng, R.; Lai, Y.; Liu, Y.; Wang, M. Breaking the limitation of polarization multiplexing in optical metasurfaces with engineered noise, *Science*, **2023**, *379*, 294-299.

(5) Xing, C.; Zhou, B.; Yan, D.; Fang, W. Integrating full-color 2D optical waveguide and heterojunction engineering in halide microsheets for multichannel photonic logical gates, *Adv. Sci*. **2024**, *11*, 2310262.

(6) Lee, J.; Bisso, P. W.; Srinivas, R. L.; Kim, J. J.; Swiston, A. J.; Doyle, P. S. Universal process-inert encoding architecture for polymer microparticles, *Nat. Mater.* **2014**, *13*, 524.

(7) Zijlstra, P.; Chon, J. W. M.; Gu, M. Five-dimensional optical recording mediated by surface plasmons in gold nanorods, *Nature* **2009**, *459*, 410.

(8) Chen, T.; Yan, D. Full-color, time-valve controllable and Janus-type long-persistent luminescence from all-inorganic halide perovskites, *Nat. Commun.* **2024**, *15*, 5281.

(9) Xiao, G.; Zhou, B.; Fang, X.; Yan, D. Room-temperature phosphorescent organic-doped inorganic frameworks showing wide-range and multicolor long persistent luminescence, *Research* **2021**, *2021*, 9862327.

(10) Riesen, H.; Liu. Z. Optical storage phosphors and materials for ionizing radiation, *In Current Topics in Ionizing Radiation Research* **2012**, *29*, 625.

(11) Zhuang, Y.; Chen, D.; Chen, W.; Zhang, W.; Su, X.; Deng, R.; An, Z.; Chen, H.; Xie, R. X-ray-charged bright persistent luminescence in $NaYF_4:Ln^{3+}@NaYF_4$ nanoparticles for multidimensional optical information storage, *Light Sci. Appl.* **2021**, *10*, 132.

(12) Liu, D.; Yuan, L.; Wu, Y. H.; Lv, Y.; Xiong, G.; Ju, G.; Chen, L.; Yang, S.; Hu, Y. Tailoring multidimensional traps for rewritable multilevel optical data storage, *ACS Appl. Mater. Interfaces* **2019**, *11*, 35023.

(13) Long, Z.; Zhou, J.; Qiu, J.; Wang, Q.; Li, Y.; Wang, J.; Zhou, D.; Yang, Y.; Wu, H.; Wen, Y. Thermal engineering of electron-trapping materials for "Smart-Write-In" optical data storage, *Chem. Eng. J.* **2021**, *420*, 129788.

(14) Lin, S.; Lin, H.; Ma, C.; Cheng, Y.; Ye, S.; Lin, F.; Li, R.; Xu, J.; Wang, Y. High-security-level multi-dimensional optical storage medium: nanostructured glass embedded with $LiGa_5O_8:Mn^{2+}$ with photostimulated luminescence, *Light Sci. Appl.* **2020**, *9*, 22.

(15) Tang, H.; Liu, Z.; Zhang, H.; Zhu, X.; Peng, Q.; Wang, W.; Ji, T.; Zhang, P.; Le, Y.; Yakovlev, A. N.; Qiu, J.; Dong, G.; Yu, X.; Xu, X. 4D optical information storage from $LiGa_5O_8:Cr^{3+}$ nanocrystal in glass, *Adv. Opt. Mater.* **2023**, *11*, 2300445.

(16) Xing, C.; Zhou, B.; Yan. D.; Fang. W. Dynamic photoresponsive ultralong phosphorescence from one-dimensional halide microrods toward multilevel information storage, *CCS Chemistry*, **2023**, *5*, 2866.

(17) Pan, Z.; Lu, Y.; Liu. F. Sunlight-activated long-persistent luminescence in the near-infrared from $Cr^{3+}$-doped zinc gallogermanates, *Nat. Mater.* **2012**, *11*, 58.

(18) Zhou, B.; Yan, D. Glassy inorganic-organic hybrid materials for photonic applications, *Matter* **2024**, *7*, 1950-1976.

(19) Xing, C.; Qi, Z.; Zhou, B.; Yan, D.; Fang, W. Solid-state photochemical cascade process boosting smart ultralong room-temperature phosphorescence in bismuth halides, *Angew.Chem.* **2024**, *163*, e202402634.

(20) Li, Y.; Gecevicius, M.; Qiu. J. Long persistent phosphors—from fundamentals to applications, *Chem. Soc. Rev.* **2016**, *45*, 2090.

(21) Matsuzawa, T.; Aoki, Y.; Takeuchi, N.; Murayama, Y. A new long phosphorescent phosphor with high brightness, $SrAl_2O_4$:$Eu^{2+}$,$Dy^{3+}$, *J. Electrochem. Soc.* **1996**, *143*, 2670.

(22) Dorenbos, P. Mechanism of persistent luminescence in $Sr_2MgSi_2O_7$:$Eu^{2+}$; $Dy^{3+}$, *Phys. Stat. Sol (b)* **2005**, *242*, R7.

(23) Kresse, G.; Furthmüller, J. Efficiency of ab–initio total energy calculations for metals and semiconductors using a plane–wave basis set, *Comput. Mater. Sci.* **1996**, *6*, 15.

(24) Kresse, G.; Joubert, D. From ultrasoft pseudopotentials to the projector augmented-wave method, *Phys. Rev. B* **1999**, *59*, 1758.

(25) Perdew, J. P.; Burke, K.; Ernzerhof, M. Generalized gradient approximation made simple, *Phys. Rev. Lett.* **1996**, *77*, 3865.

(26) Dudarev, S. L.; Botton, G. A.; Savrasov, S. Y.; Humphreys, C. J.; Sutton, A. P. Electron-energy-loss spectra and the structural stability of nickel oxide: An LSDA+ U study, *Phys. Rev. B* **1998**, *57*, 1505.

(27) Ou, X.; Qin, X.; Huang, B.; Zan, J.; Wu, Q.; Hong, Z.; Xie, L.; Bian, H.; Yi, Z.; Chen, X.; Wu, Y.; Song, X.; Li, J.; Chen, Q.; Yang, H.; Liu, X. High-resolution X-ray luminescence extension imaging, *Nature* **2021**, *590*, 410.

(28) Luo, H.; Dorenbos, P. The dual role of $Cr^{3+}$ in trapping holes and electrons in lanthanide co-doped $GdAlO_3$ and $LaAlO_3$, *J. Mater. Chem. C* **2018**, *6*, 4977.

(29) Wang, X.; Chen, Y.; Kner, P. A.; Pan, Z. $Gd^{3+}$-activated narrowband ultraviolet-B persistent luminescence through persistent energy transfer, *Dalton Trans.* **2021**, *50*, 3499.

(30) Meng, J.; Zhang, L.; Yao, F.; Liu, X.; Meng, J.; Zhang, H. Density Functional characterization of the 4f-relevant electronic transitions of lanthanide-doped $Lu_2O_3$ luminescence materials, *Phys. Chem. Chem. Phys.* **2018**, *19*, 2947.

(31) Wang, F.; Deng, R.; Wang, J.; Wang, Q.; Han, Y.; Zhu, H.; Chen, X.; Liu, X. Tuning upconversion through energy migration in core–shell nanoparticles, *Nat. Mater.* **2011**, *10*, 968.

(32) Liang, Y.; Liu, F.; Chen, Y.; Sun, K.; Pan, Z. Long persistent luminescence in the ultraviolet in $Pb^{2+}$-doped $Sr_2MgGe_2O_7$ persistent phosphor, *Dalton Trans.* **2016**, *45*, 1322.

(33) Zhang, S.; Zhao, F.; Liu, S.; Song, Z.; Liu, Q. An improved method to evaluate trap depth from thermoluminescence, *J. Rare Earths* 2024, https://doi.org/10.1016/j.jre.2024.02.004.

(34) He, Z.; Wang, X.; Yen, W. Investigation on charging processes and phosphorescent efficiency of $SrAl_2O_4$:Eu,Dy, *J. Lumin.* **2006**, *119*, 309.

(35) Shionoya, S.; Kallmann, H. P.; Kramer, B. Behavior of excited electrons and holes in zinc sulfide phosphors, *Phys. Rev.* **1961**, *121*, 1607.

(36) Deng, M.; Liu, Q.; Zhang, Y.; Wang, C.; Guo, X.; Zhou, Z.; Xu, X. Novel co-doped $Y_2GeO_5$:$Pr^{3+}$,$Tb^{3+}$: Deep trap level formation and analog binary optical storage with submicron information points, *Adv. Opt. Mater.* **2021**, *9*, 2002090.

(37) Liang, L.; Chen, J.; Shao, K.; Qin, X.; Pan, Z.; Liu, X. Controlling persistent luminescence in nanocrystalline phosphors, *Nat. Mater.* **2023**, *22*, 289.

(38) Joos, J. J.; Korthout, K.; Amidani, L.; Glatzel, P.; Poelman, D.; Smet, P. F. Identification of $Dy^{3+}/Dy^{2+}$ as electron trap in persistent phosphors, *Phys. Rev. Lett.* **2020**, *125*, 033001.

(39) Peng, F.; Seto, T.; Wang, Y. First Evidence of Electron Trapped $Re^{2+}$ Promoting Afterglow on $Eu^{2+}$, $Ln^{3+}$ Activated Persistent Phosphor-Example of $BaZrSi_3O_9$:$Eu^{2+}$, $Sm^{3+}$, *Adv. Funct. Mater.* **2023**, *33*, 2300721.

(40) Zhou, X.; Ning, L.; Qiao, J.; Zhao, Y.; Xiong, P.; Xia, Z. Interplay of defect levels and rare earth emission centers in multimode luminescent phosphors, *Nat. Commun.* **2022**, *13*, 7589.

(41) Huang, K.; Le, N.; Wang, J.; Huang, L.; Zeng, L.; Xu, W.; Xu, Z.; Li, Y.; Han, G. Designing next generation of persistent luminescence: recent advances in uniform persistent luminescence nanoparticles, *Adv. Mater.* **2022**, *34*, 2107962.

(42) Li, S.; Zhu, Q.; Xiahou, J.; Li, J. Polyhedron engineering by chemical unit cosubstitution in $LaAlO_3$:$0.02Pb^{2+}$ to generate multimode and condition-sensitive luminescence for dynamic anticounterfeiting, *Chem. Eng. J.* **2022**, *450*, 138440.

(43) Bai, X.; Yang, Z.; Zhan, Y.; Hu, Z.; Ren, Y.; Li, M.; Xu, Z.; Ullah, A.; Khan, I.; Qiu, J.; Song, Z.; Liu, B.; Wang, Y. Novel strategy for designing photochromic ceramic: reversible upconversion luminescence modification and optical information storage application in the $PbWO_4$: $Yb^{3+}$, $Er^{3+}$ photochromic ceramic, *ACS Appl. Mater. Interfaces* **2020**, *12*, 21936.

(44) Wu, L.; Sun, S.; Bai, Y.; Xia, Z.; Wu, L.; Chen, H.; Zheng, L.; Yi, H.; Sun, T.; Kong, Y.; Zhang, Y.; Xu, J. Defect-induced self-reduction and anti-thermal quenching in $NaZn(PO_3)_3$:$Mn^{2+}$ red phosphor, *Adv. Opt. Mater.* **2021**, *9*, 2100870.

(45) Zhang, S.; Hu, Y.; Chen, L.; Wang, X.; Ju, G.; Wang, Z. Systematic investigation of photoluminescence on the mixed valence of europium in $Zn_2GeO_4$ host, *Appl. Phys. A* **2014**, *116*, 1985.

(46) Lyu, T.; Dorenbos, P. $Bi^{3+}$ acting both as an electron and as a hole trap in La-, Y-, and $LuPO_4$, *J. Mater. Chem. C* **2018**, *6*, 6240.

(47) Van der Heggen, D.; Joos, J. J.; Feng, A.; Fritz, V.; Delgado, T.; Gartmann, N.; Walfort, B.; Rytz, D.; Hagemann, H.; Poelman, D.; Viana, B.; Smet, P. F. Persistent luminescence in strontium aluminate: a roadmap to a brighter future, *Adv. Funct. Mater.* **2022**, *32*, 2208809.

(48) Bian, H.; Qin, X.; Wu, Y.; Yi, Z.; Liu, S.; Wang, Y.; Brites, C. D. S.; Carlos, L. D.; Liu, X. Multimodal tuning of synaptic plasticity using persistent luminescent memitters, *Adv. Mater.* **2022**, *34*, 2101895.

(49) Du, J.; Wang, X.; Sun, S.; Wu, Y.; Jiang, K.; Li, S.; Lin, H. Pushing trap-controlled persistent luminescence materials toward multi-responsive smart platforms: recent advances, mechanism, and frontier applications. *Adv. Mater.* **2024**, *36*, 2314083.

**Graphic Abstract**

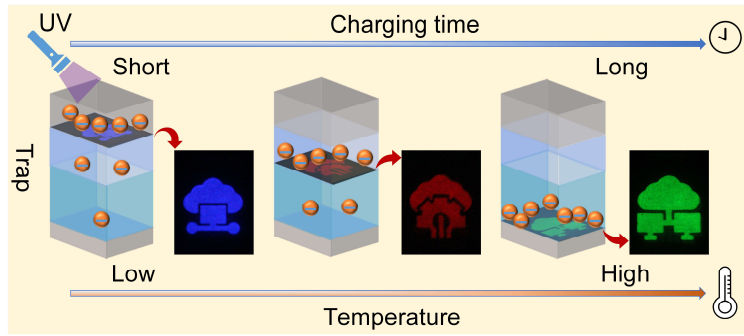